\documentclass[pra,onecolumn,aps]{revtex4}

\usepackage{amssymb}
\usepackage{amsmath}
\usepackage[dvips]{graphicx}

\renewcommand{\deg }{$^{\circ}$}

\begin{document}

\title{Potential Phytoextraction with in-vitro regenerated plantlets 
of Brassica juncea (L.) Czern. in presence of CdCl$_2$: 
Cadmium accumulation and physiological parameter measurement.}

\author{ Michel AOUN\S}\email{ michel.aoun@univ-brest.fr}
\author{Jean-Yves CABON\P, Annick HOURMANT\S}
 \affiliation{\S Laboratoire de Biotechnologie et Physiologie V\'eg\'etales, 
Universit\'e de Bretagne Occidentale (Brest - France). \\
\P Laboratoire de Chimie, Electrochimie Mol\'eculaire et Chimie Analytique - 
UMR 6521, Universit\'e de Bretagne Occidentale (Brest - France).}

\begin{abstract}
Heavy metal contamination of agricultural land  is partly responsible for 
limiting crop productivity. 
Cd$^{2+}$ is known as a non-essentiel HM that can be harmful to plants even at low 
concentrations. \\
Brassica juncea (L.) is able to accumulate more than 400 $\mu$g.g$^{-1}$ D.W. in the 
shoot, a physiological trait which may be exploited for the phytoremediation of 
contaminated soils and waters. \\
In our study, we have subjected regenerated plantlets obtained by the means of 
transverse thin cell layer technology from hypocotyl of B. juncea in the 
presence of 150 $\mu$M CdCl$_2$ to a second stress of CdCl$_2$ (75 $\mu$M), applied 
hydroponically for 3 days, to show whether the in-vitro plantlets were able to 
accumulate more Cd than wild type (plantlets obtained in the absence of 150 $\mu$M 
CdCl$_2$, and hence argue about their potential use in the phytoextraction process. The 
application of 75 $\mu$M CdCl$_2$ for three days does not show any effect in the B. 
juncea growth parameters (F.W. and D.W.) whatever the type of plantlets. This 
application decreases also the contents of chlorophyll a, carotenoids and Chl a/b 
ratio (2.26) for plantlets regenerated in the absence of CdCl$_2$ but not 
those of plantlets regenerated in its presence. Moreover, the amounts of MDA 
were increased in all plantlets however even more in those obtained in the presence of 
CdCl$_2$. Cd contents were higher in the plantlets regenerated in the absence of 
75 $\mu$M CdCl$_2$ than those in its presence. Roots have the highest contents (3071; 
1544 $\mu$g.g$^{-1}$ D.W.) followed by stems (850; 687$\mu$g.g$^{-1}$ D.W.) and leaves (463; 
264$\mu$g.g$^{-1}$ D.W.) respectively. \\
In our conditions, we suggest that the low accumulation in the plantlets 
regenerated in the presence of CdCl$_2$ by the means of in-vitro regeneration 
technology is still benefical, to some extent, for the phytoextraction process and 
seems to be an interesting technology that allows the cultivation of these 
plantlets in contaminated soils with low accumulation of metal in 
their shoots and probably in their seeds used in many food technologies. This results in  
decreasing the health risk du to the consumption of the spinoff products.

\end{abstract}

\maketitle


\textbf{Abbreviations} D.W. Dry weight, F.W. Fresh weight, Heavy metals  HM, MDA Malondialdehyde,
MES 2-(N-morpholino ethane sulfonic) acid, MS Murashige and Skoog medium (1962), tTCL(s) 
transverse Thin Cell Layer(s), S.D.  Standard Deviation.

\section{Introduction}

Heavy metal (HM) contamination of soils has become a serious problem in areas of 
intense industry and agriculture. Increasing of HM concentrations in polluted 
soils pose a serious health hazard to man, animals, plants as well as soil 
micro-organisms, vitally important for soil health and fertility, especially those 
which are  particularly sensitive to HM stress (Dahlin et al. 1997) and their biological 
diversity in the soil is reduced by HM contamination (Giller et al. 
1998).\\

The remediation of contaminated soils by conventional technologies (e.g. 
excavation, physical stabilization or washing) used for small areas of heavily 
contaminated soils is not economically feasible. Phytoextraction, the use of 
plants for extraction of HM from contaminated sites, has emerged as 
an economically viable and socially acceptable decontamination for remediation 
of low to medium polluted soil (Salt et al. 1995a).\\

Cadmium is an HM  naturally occuring in soils at low 
concentration (Traina, 1999). It is a toxic metal that can be harmful to 
animals and plants. Plants are able to absorb Cd via roots, to translocate it 
to leaves and thus introduce it into the food chain (Hart et al. 1998; Cakmak et 
al. 2000). \\

In the recent past decades, Brassica juncea has attracted 
researchers because of its high biomass production with added economical value 
and its high capacity to translocate and accumulate many metals and metalloids 
as As, Cd, Cu, Pb, Se, and Zn from polluted soils and therefore, can be 
considered as a good candidate in phytoextraction process (Kumar et al. 
1995; Salt et al. 1995b; Blaylock et al. 1997; Raskin et al. 1997).\\

The aim of our study is to investigate the potential use of B. juncea plantlets 
regenerated from hypocotyl tTCl explants in the presence of CdCl$_2$ in the 
phytoextraction process. \\

\section{Material and methods}

\textbf{Plant material} \\

Brassica juncea AB79/1 was used in our study. This cultivar is pure spring 
line, genetically fixed and was obtained by autofertilization. \\

\textbf{Culture condition and regeneration of plants} \\

Seeds of Brassica juncea were decontaminated in 70\% ethanol for 30 sec, 
followed by immersion in calcium hypochlorite (5\%, w/v) added with two drops of 
Tween-20 for 10 min. The seeds were rinsed twice for 5 min with sterile water 
upon sterilization and sown in test tubes on MS medium containing sucrose 2 \% 
(w/v) and solidified with agar at 6 g.l$^{-1}$ (Kalys, HP 696). They were incubated 
later on under a photoperiod of 12h (60 $\mu$mol photon.m$^{-2}$s$^{-1}$) provided by cool 
white fluorescent lamps, with a 22/20\deg C thermoperiod (light/dark). \\

tTCLs (400-500 $\mu$m) were excised from hypocotyls of 7 day-old Brassica juncea 
seedlings and cultivated in a Petri dishes containing MS medium (25 ml) (15 
tTCLs per Petri dish). \\

MS medium (comprising macronutriments, micronutriments and vitamins of 
Murashige and Skoog, 1962) supplemented with BAP (26.6 $\mu$M), NAA (3.22 $\mu$M), 
Sucrose (20 g.l$^{-1}$), AgNO$_3$ (10 $\mu$M) and CdCl$_2$ (0 or 150 $\mu$M). All media were 
solidified with agar (0.6 \%, w/v), adjusted to pH 5.8 by 0.1 N NaOH and 
sterilized by autoclaving at 120\deg C for 20 min. \\

All cultures were incubated in the same conditions as previously described. 
After 6 weeks, shoots were separated and transferred to test tubes containing 
MS medium (10 ml) without either, CdCl$_2$ and PGRs to induce rooting. The small 
plantlets were transferred to pots containing sterile vermiculite (EFISOL, 
VERMEX M) in a naturally-lighted greenhouse, watered daily and fertilized with 
half strength Hoagland solution (Hoagland and Arnon, 1950). \\

\textbf{Hydroponic application of CdCl$_2$} \\

Indian mustard (Brassica juncea) plantlets were grown on humidified vermiculite 
at 22\deg C. When seedlings have 18 days, they are extracted from vermiculite and 
then roots are washed carefuly under water flow. Then, 75 $\mu$M CdCl$_2$ was applied 
hydroponically in a half strength Hoagland solution (Hoagland and Arnon, 1950) 
for 3 days in presence of MES buffer (1 mM). All experiments were conducted in a 
growth chamber under a photoperiod of 16h (60 $\mu$mol photon.m$^{-2}$s$^{-1}$) 
provided by cool white fluorescent lamps, with a 22/20\deg C thermoperiod (light/dark). The 
final pH in the all culture medium, was 6.60 $\pm$ 0.05. \\

\textbf{Determination of pigment contents} \\

The contents of chlorophyll a and b and carotenoids are determined by 
Lichtenthaler procedure (Lichtenthaler, 1987). Approximatively 1g of limb 
fresh weight is extracted by acetone 100 \% (25 ml), and then the mixture is 
centrifuged at 5000 rpm for 10 min before reading the wavelength at 470, 662 
and 645nm respectively using a Shimadzu UV-visible spectrophotometer (UV-1605). \\

\textbf{Measurement of Cd in the different organs} \\

The determination of Cd concentrations in the different digested solutions 
is conducted with electrothermal atomic absorption spectrometry. A Perkin-Elmer 
SIMAA 6100 working in single element monochromator mode was used for all 
atomic absorption measurements. At harvest (84 h of treatment), leaves, stems 
were weighed and then oven-dried for 4 days at 80\deg C, however, roots were 
extensively washed in distilled water for 10 min, then weighed before 
oven-dried for 4 days at 80\deg C. For the preparation of all solutions, high- 
purity water from a MilliQ-system (Millipore, Milford, MA, USA) was used. 
Sample aliquots of approximately 200 mg were transferred into the Teflon 
vessels. After addition of acid mixture: nitric acid, hydrogen peroxide and 
hydrofluoric acid to the powders in the ratios (4/3/1, v/v/v), the vessels were 
closed and exposed to microwaves digestion as described in detail elsewhere 
(Weiss et al. 1999). \\

\textbf{Estimation of lipid peroxidation} \\

The level of lipid peroxidation was determined as malondialdehyde (MDA) content 
able te react with thiobarbituric acid and was measured according to Minotti 
and Aust (1987) and Iturbe-Ormaetxe et al. (1998). Approximatively 1 g limb 
fresh weight was extracted by 6 ml of a mixture of meta-phosphoric acid (5 \%, 
w/v) and butylhydroxy-toluene (50/1, v/v) the homogenate was centrifuged at 
5000 rpm for 30 min and then 4 ml of supernatant were homogenized with 200 $\mu$L 
of butyhydroxytoluene (2 \%, w/v), 1 mL of HCl (25 \%, v/v) and 1 mL of 
2-thiobarbituric acid (1 \%, w/v) prepared in NaOH (50 mM). The homogenate was 
incubated at 95\deg C for 30 min followed by rapid cooling in an ice bath to stop 
the reaction. Finally, we added 3 ml of 1-butanol on the homogenate then 
centrifuged at 500 rpm for 5 min before reading the absorbance at 532 nm. MDA 
contents were calculated using the extinction coefficient of MDA ($\alpha$ = 155 mmol 
L$^{-1}$ cm$^{-1}$). \\

\textbf{Data analysis} \\

Each experiment was repeated 3 times with 3 independent runs. For all 
parameters,The values were compared by analysis of variance (ANOVA) and the 
differences among means (5\% level of significance) were tested by the LSD test 
using StatGraphics Plus 5.1. \\

\section{Results}

\textbf{Cadmium accumulation in B. juncea regenerted plantlets} \\

Regenerated plantlets, from hypocotyl tTCLs cultivated in the presence of 150 
$\mu$M CdCl$_2$ for 6 weeks, showed an accumulation of 154.8 and 46.3 $\mu$g Cd.g$^{-1}$ D.W. 
in their root and stem tissues respectively, but not in their leaves (Table 1).
When the plantlets were subjected to 75 $\mu$M CdCl$_2$ for 3 days, the accumulation 
of Cd ($\mu$g.g$^{-1}$ D.W.) was enhanced in the order : Roots (3070.84) $>>$ stem 
(849.34) $>$ Leaves (463.17). However, plantlets regenerated from hypocotyl tTCLs 
in the presence of 150 $\mu$M CdCl$_2$ showed a decrease of the Cd accumulation when 
they were subjected to 75 $\mu$M CdCl$_2$; indeed, the accumulation of Cd decrease 
1.98 and 1.75 in root and leaf tissues respectively (Table 1). \\

\textbf{Effect of CdCl$_2$ on  growth} \\

The presence of 150 $\mu$M CdCl$_2$ in the culture medium of tTCLs induces the 
inhibition of growth of neoformed buds; this inhibition was observed in 
plantlets after their transfer to growth chamber for 5 weeks in the absence of 
CdCl$_2$ and resulting from a significant decrease of fresh and dry weight of 
leaves (56 and 48 \%), stem (64 and 66 \%) and roots (54 and 61 \%).
However, when regeneratd plantlets in the presence or not of 150 $\mu$M CdCl$_2$ were 
subjected hydroponically to 75 $\mu$M CdCl$_2$ for 3 days, their growth (fresh and dry 
weight) was not significantly modified per comparison with their respective 
plantlet controls (Table 2 and 3). \\

\textbf{Effect of CdCl$_2$ on the pigment contents} \\

The plantlets regenerated from hypocotyl tTCLs in the presence or absence of 150 $\mu$M 
CdCl$_2$ did not show any significant changes in their pigment contents (Figure 1) 
and the Chl a / b ratios were 2.95 for control and 3.24.
However, the application of 75 $\mu$M CdCl$_2$ did not decrease pigment contents of 
plantlets regenerated in the presence of 150 $\mu$M CdCl$_2$, but decrease the Chl a 
and carotenoid contents by 33 \% of plantlets regenrated in its presence per 
comparison to control plants. we observed also the decrease of Chl a / b ratio 
to the value of 2.26. Moreover, whatever the treatment, Chl b contents did not 
produce any significant effect (Figure 1). \\

\textbf{Effect of CdCl$_2$ on lipid peroxidation and malondialdehyde (MDA) levels} \\

The levels of malondialdehyde was used as an indicator of lipid peroxidation 
and they were measured at leaf levels.
The application of 75 $\mu$M CdCl$_2$ for 3 days on plantlets regnerated in the 
absence of CdCl$_2$ enhanced significantly the MDA formation (+ 55 \%) in 
comparison to control plants (figure 2). \\

The regeneration of plantlets in the presence of 150 $\mu$M CdCl$_2$ did not show any 
significant difference in the lipoperoxid production in comparison to control, 
but this one was enhanced significantly after 3 days application of 75 $\mu$M 
CdCl$_2$; indeed, the MDA production was 2.82 fold more than the conrol plants 
(Figure 2). \\

\section{Discussion and Conclusions}

Brassica juncea has been identified as a high biomass-producing crop with the 
capacity to take up and accumulate many HM and metalloids (Salt et 
al. 1995b; Blaylock et al. 1997; Raskin et al. 1997). This study on B. juncea 
provides the first comprehensive analysis of Cd accumulation in in-vitro 
regenerated plants. \\

Significant differences were found between the type of regenerant. Indeed, 
plantlets regenerated in the presence of 150 $\mu$M CdCl$_2$ accumulated less Cd than 
plantlets regenerated in its absence when all these plantlets were subjected to 
75 $\mu$M CdCl$_2$ applied hydroponically. This decrease was observed for all organ of 
plantlets (Table 1). \\

Intrestingly, the maximum Cd concentration found in our study (463.17 $\mu$g Cd.g$^{-1}$ 
D.W.) was similar to the concentration found in B. juncea plants exposed to Cd 
for the same time exposure (Haag-Kerwer et al. 1999) and to much lower solution 
concentrations (Salt et al. 1995b, 1997; Haag-Kerwer et al. 1999), indicating 
an overall limitation of Cd uptake by the leaf tissues in B. juncea plants, 
irrespective of the Cd concentration administered to the plant in the medium. 
Possible reasons could be limitations of root to shoot translocation, and/or a 
saturable capacity for intracellular detoxification of Cd (Haag-Kerwer et al. 
1999).  \\

In our result, this partial exclusion of Cd seems to result more probably from 
the inhibition of transporters or ionic channels implicated in the root 
absorption as demonstrated by Cseh (2002) and Ann Peer et al. (2003) or to a 
change in the capacity of cell wall to bind metal or to an enhanced excretion 
of chelated substances as discussed in excluder plants by Ghosh and Singh 
(2005) and Kirkham (2006). Indeed, whether the Cd amount in roots were reduced 
by 50 \% per comparison to the alone application of 75 $\mu$M CdCl$_2$, the amounts of 
aerial parts were reduced only by 28 \%. \\

The Cd-stress (150 $\mu$M) was found to adversely affect plant growth of 
regenerated plantlets but not 75 $\mu$M CdCl$_2$ applied hydroponically for 3 days. 
Our results show that the presence of 150 $\mu$M CdCl$_2$ in the culture medium of 
hypocotyl tTCL explants induced the growth reduction of neoformed plantlets 
(Table 2). \\

Indeed, although the plantlets regenerated in the presence of 150 $\mu$M CdCl$_2$ were 
cultivated after, with absence of this metal for 5 weeks, they showed a 
reduction of fresh weight of leaves (56 \%), stem (64 \%) and roots (53 \%). At 
the same time, the dry weight of all organ was reduced, with the less observed 
effect at foliar level and the most one at root level (Table 3). \\

Growth reduction in response to Cd-stress was also reported for many species 
such as bean (Poschenrieder et al. 1989), willow, poplar (Lunackova et al. 
2003; Cosio et al. 2005), rice (Aina et al. 2007), sunflower (Groppa et al. 
2007), and some Brassica species such as Brassica napus (Larsson et al. 1998) 
and Brassica juncea (Haag-Kerwer et al. 1999); Cd inhibition was function of 
time exposure and concentration of Cd. \\

The second application of 75 $\mu$M CdCl$_2$ hydroponically did not affect 
significantly the growth parameters (F.W. and D.W.) of different organ of 
plantlets regenerated from hypocotyl tTCL explants.
Chlorophyll contents were usually used to evaluate the impact of many 
environmental stresses on plant health. In Our investigation, we showed that 
the regeneration in the presence of 150 $\mu$M CdCl$_2$ did not modify the chlorophyll 
and carotenoid contents in 5 week-old neoformed plantlets. It seems that the 
growth retardation due to the Cd presence in the culture medium of tTCL 
explants and which persists for 5 weeks did not affect the biosynthesis of 
pigments. \\

However, the hydroponically application of 75 $\mu$M CdCl$_2$ induced a reduction of 
chlorophyll a and carotenoid contents according to many works that reported a 
decrease of chlrophyll contents under Cd-stress (Padmaja et al. 1990; Larsson 
et al. 1998; Groppa et al. 2007a, b) but also in the presence of other metals 
(Chatterjee and Chatterjee, 2000; Mysliwa-Kurdziel and Strzalka, 2002; Lei et 
al. 2007). The decrease of chlorophyll contents was one of primary events in 
plants subjected to metal stress and resulted from the inhibition of enzymes 
responsible to the biosynthesis of pigments (Stobart et al. 1985; 
Mysliwa-Kurdziel and Strzalka, 2002). As described by Mysliwa-Kurdziel and 
Strzalka (2002), Cd affected 2 pathways in this biosynthesis: it inhibited the 
aminolevulinic acid synthesis and the redcution of protochlorophyllid to 
chlorophyllid. \\

Moreover, hydroponically applied Cd in B. juncea plants in our study, induced a 
decrease of Chl a / b ratio agreed with the investigations of Baszynski et al. 
(1980) and Larsson et al. (1998). The Chl b contents being not affected, it is 
likely that reduction of Chl a contents may result from more sensitivity to Cd 
toxicity and then weaker synthesis or a faster degradation of this 
pigment.  \\

The Cd-stress induce also a reduction of carotenoid contents agreeing 
with many investigations for many plant species (Baszynski et al. 1980; Larsson 
et al. 1998; Mysliwa-Kurdziel and Strzalka, 2002). The significant accumulation 
of Cd in leaves, probably responsible of free radical production, testified 
by the significant production of MDA. Indeed, this production may lead to the 
destruction of Chl a and the antioxidant carotenoids. \\

The application of 75 $\mu$M CdCl$_2$ hydroponically shows an enhanced production of 
MDA and an enhanced lipid peroxidation. It is known that the 
peroxidation of polyinsaturated fatty acids of cellular membrane disturb their 
functions, especially by the reduction of membrane fluidity and the 
inhibition of receptors and enzymes located in the cellular membrane (Lagadic 
et al. 1997). \\

Summarizing our results, we conclude that plantlets regenerated in the 
presence of CdCl$_2$ by the means of tTCL technology while benefical, to some 
extent, for the phytoextraction process seems to be an interesting 
technology for the cultivation of plantlets in a contaminated 
soils with a low accumulation ability of HM in their tissues. Ultimately, 
this decreases health risks due to the consumption of their derivative products.

\vspace{1cm}

\textbf{Acknowledgement}

We thank Dr Thierry Guinet from Ecole Nationale d'Enseignement Sup\'erieure 
Agronomique (ENESA) de Dijon (France)' for furnishing the seeds of spring line 
AB79/1 of Brassica juncea.

\vspace{1cm} 

\textbf{References} \\

Aina, R., Labra, M., Fumagalli, P., Vannini, C., Marsoni, M., Cucchi, U., 
Bracale, M., Sgorbati, S., Citterio, S., 2007. Thiol-peptide level and 
proteomic changes in response to cadmium toxicity in Oryza sativa L. roots. 
Environ. Exp. Bot. 59, 381-392.

Baszynski, T., Wajda, L., Krol, M., Wolinska, D., Krupa, Z., Tukendorf, A., 
1980. Photosynthetic activities of cadmium-treated tomato plants. Physiol. 
Plant. 48, 365-370.

Blaylock, M.J., Salt, D.E., Dushenkov, S., Zakharova, O., Gussman, C., 
Kapulnik, Y., Ensley, B.D., Raskin, I., 1997. Enhanced accumulation of Pb in 
Indian mustard by soil-applied chelating agents. Environ. Sci. Technol. 31, 
860-865.

Cakmak, I., Welch, R.M., Hart, J., Norwell, W.A., Ozturk, L., Kochian, L.V., 
2000. Uptake and translocation of leaf-applied cadmium in diploid, tetraploid 
and hexaploid wheats. J. Exp. Bot. 51, 221-226.

Chatterjee, J., Chatterjee, C., 2000. Phototoxicity of cobalt, chromium and 
copper in cauliflower. Environ. Pollut. 109, 69-74.

Cosio, C., Vollenweider, P., Keller, C., 2005. Localization and effects of 
cadmium in leaves of a cadmium-tolerant willow (Salix viminalis L.). I. 
Macrolocalization and phytotoxic effects of cadmium. Environ. Exp. Bot. 

Dahlin, S., Witter, E., Martensson, A., Turner, A., Baath, A., 1997. Where's 
the limit? Changes in the microbiological properties of agricultural soils at 
low levels of metal contamination. Soil Biol. Biochem. 29, 1405-1415.

Giller, K.E., Witter, E., McGrath, S.P., 1998. Toxicity of HMs to 
microorganisms and microbial processes in agricultural soils: A review. Soil 
Biol. Biochem. 30, 1389-1414.

Ghosh, M., Singh, S.P., 2005. A review on phytoremediation of HMs and 
utilization of its byproducts. Appl. Ecol. Environ. Res. 3, 1-18.

Groppa, M.D., Ianuzzo, M.P., Tomaro, M.L., Benavides, M.P., 2007a. Polyamine 
metabolism in sunflower plants under long-terme cadmium or copper stress. Amino 
Acids 32, 265-275.

Groppa, M.D., Tomaro, M.L., Benavides, M.P., 2007b. Polyamines and HM 
stress : the antioxidant behavior of spermine in cadmium- and copper-treated 
wheat leaves. BioMetals 20, 185-195.

Haag-Kerwer, A., Schäfer, H.J., Heiss, S., Walter, C., Rausch, T., 1999. 
Cadmium exposure in Brassica juncea causes a decline in transpiration rate and 
leaf expansion without effect on photosynthesis. J. Exp. Bot. 50, 1827-1835.

Hart, J.J., Welch, R.M., Norvell, W.A., Sullivan, L.A., Kochian, L.V., 1998. 
Characterization of cadmium binding, uptake, and translocation in intact 
seedlings of bread and durum wheat cultivars. Plant Physiol. 116, 1413-1420.

Hoagland, D.R., Arnon, D.I., 1950. The water-culture for growing plants without 
soil. Cal. Agric. Exp. Sta Cir. 347, 1-32.

Iturbe-Ormaetxe, I., Escuredo, P.R., Arrese-Igor, C., Becana, M., 1998. 
Oxidative damage in pea plants exposed to water deficit or paraquat. Plant 
Physiol. 116: 173-181.

Kirkham, M.B., 2006. Cadmium in plants on polluted soils : Effects of soil 
factors, hyperaccumulation, and amendments. Geoderma 137, 19-32.

Kumar, P.B.A.N., Dushenkov, V., Motto, H., Raskin, I., 1995. Phytoextraction : 
The use of plants to remove HMs from soils. Environ. Sci. Technol. 29, 
1232-1238.

Lagadic, L., Caquet, T., Amiard, J.-C., Ramade, F., 1997. M\'ecanismes de 
formation et effets des esp\`eces r\'eactives de l'oxyg\`ene. In : Biomarqueurs en 
\'ecotoxicologie : Aspects fondamentaux , pp, 125-147.

Larsson, E.H., Bornman, J.F., Asp, H., 1998. Influence of UV-radiation and Cd$^{2+}$ 
on chlorophyll fluorescence, growth and nutrient content in Brassica napus. J. 
Exp. Bot. 323, 1031-1039.

Lei, Y.B., Korpelainen, H., Li, C.Y., 2007. Physiological and biochemical 
responses to high Mn concentrations in two contrasting Populus cathayana 
populations. Chem. 68, 686-694.

Lichtenthaler, H.K., 1987. Chlorophylls and carotenoids: Pigments of 
photosynthetic biomembranes. Methods Enzymol. 148, 351-382.

Lunackova, L., Sottnikova, A., Masarovicova, E., Lux, A., Stresko, V., 2003, 
Comparison of cadmium effect on willow and poplar in response to different 
cultivation conditions. Biol. Plant. 47, 403-411.

MacAdam, J.W., Nelson, C.J., Sharpe, R.E., 1992. Peroxidase activity in the 
leaf elongation zone of tall fescue. Plant Physiol. 99, 872-878

Minotti, G., Aust, S.D., 1987. The requirement for iron (III) in the initiation 
of lipid peroxidation by iron (II) and hydrogen peroxide. J. Biol. Chem. 262, 
1098-1104.

Murashige, T., Skoog, F., 1962. Revised medium for rapid growth and bioassay 
with tobacco tissue cultures. Physiol. Plant. 15, 473-497.

Mysliwa-Kurdziel, B., Strzalka, K., 2002. Influence of metals on biosynthesis 
of photosynthetic pigments. In {\it Physiology and Biochemistry of Metal Toxicity 
and Tolerance in Plants}, eds Prasad M.N.V. et Strzalka K. pp 201-227.

Padmaja, K., Prasad, D.D.K., Prasad, A.R.K., 1990. Inhibition of chlorophyll 
synthesis in Phaseolus vulgaris L. seedlings by cadmium acetate. Phytosynth. 
24, 399-405.

Poschenreider, C., Guns\'e, B., Barcelo, J., 1989. Influence of cadmium on water 
relations, stomatal resistance and abscisic acid content in expanding bean 
leaves. Plant Physiol. 90, 1365-1371.

Raskin I., Smith, R.D., Salt, D.E., 1997. Phytoremediation of metals: using 
plants to remove polluants from the environment. Curr. Opin.Biotech. 8, 221-226.

Salt, D.E., Blaylock, M., Kumar, N., Dushenkov, V., Ensley, B.D., Chet, I., 
Raskin, I., 1995a. Phytoremediation: A novel strategy for removal of toxic 
metals from the environment using plants. Biotech. 13, 468-474.

Salt, D.E., Prince, R.C., Pickering, I.J., Raskin, I., 1995b. Mechanisms of 
cadmium mobility and accumulation in Indian mustard. Plant physiol. 109, 
1427-1433.

Salt, D.E., Pickering, I.J., Prince, R.C., Gleba, D., Dushenkov, V., Smith 
R.D., Raskin, I., 1997. Metal accumulation by aquacultured seedlings of Indian 
mustard. Environ. Sci. Technol. 3, 1636-1644.

Stobart, A.K., Griffiths, W.T., Ameen-Bukhari, I. Sherwood, R.P., 1985. The 
effect of Cd$^{2+}$ on the biosynthesis of chlorophyll in leaves of barley. Physiol. 
Plant. 63, 293-298.

Traina, S.J., 1999. The environmental chemistry of cadmium. In M.J. McLaughlin 
and B.R. Singh (ed.). Cadmium in soils and plants. Kluwer Acad. Publ.

Weiss, D.J., Shotyk, W., Schafer, J., Loyall, U., Grollimund, E., Gloor, M., 
1999. Microwave digestion of ancient peat and lead determination by 
voltammetry. Fresenius J. Anal. Chem. 363, 300-305.

\newpage

\begin{table}[htbp]
\begin{center}
\caption{Cd accumulation ($\mu$g Cd.g$^{-1}$ D.W. $\pm$ S.D.) in plantlets regenerated in the 
presence or not of CdCl$_2$ (150 $\mu$M) after the application of CdCl$_2$ (75 $\mu$M) hydroponically.} 
\label{tab1}
\begin{tabular}{ c c c c}
\hline 
CdCl$_{2}$  ($\mu $M)   & & 0  & 150   \\
\hline
 & Leaves  & 21.14 $\pm $ 1.44$^{j }$ & 20.01 $\pm $ 0.34$^{j}$  \\

0 & Stems & 25.00 $\pm $ 1.83$^{j }$  & 46.34 $\pm $ 2.90$^{h}$  \\

& Roots & 22.83 $\pm $ 2.09$^{j }$  & 154.83 $\pm $ 6.12$^{g}$  \\

\hline

& Leaves  & 463.17 $\pm $ 41.62$^{e }$ & 264.43 $\pm $ 19.03$^{f}$  \\

75 &  Stems  & 849.34 $\pm $ 56.20$^{c }$ & 686.67 $\pm $ 10.43$^{d}$  \\

& Roots  & 3070.84 $\pm $ 257.55$^{a }$ & 1544.01 $\pm $ 38.75$^{b}$  \\

\hline
\end{tabular}
\end{center}
\end{table}

 The results were calculated from three independent experiments repeated 3 times. 
For Cd levels, the values with different letters are the means $\pm$ S.D. 
and are significantly different at $p$ = 0.05 (ANOVA and LSD test).

\begin{table}[htbp]
\begin{center}
\caption{Effect of CdCl$_2$ (75 $\mu$M) hydroponically applied for 3 days 
on the growth (F.W.) of plantlets regenerated from hypocotyl tTCL explants.} 
\label{tab2}
\begin{tabular}{ c c c c}
\hline  
CdCl$_{2}$ ($\mu $M) & &  0  & 150   \\
\hline
&  Leaves  & 2670.45 $\pm $ 150.3$^{a }$ & 1170.23 $\pm $ 50.4$^{b}$  \\

0 &  Stems  & 390.34 $\pm $ 20.4$^{c }$ & 140.43 $\pm $ 10.3$^{d}$  \\

& Roots &  130.12 $\pm $ 10.12$^{d }$ & 60.23 $\pm $ 3.02$^{e}$  \\

\hline

& Leaves &  2620.23 $\pm $ 190$^{a }$ & 1165.16 $\pm $ 40.7$^{b}$  \\

75  & Stems  & 400.13 $\pm $ 20.3$^{c }$ & 142.13 $\pm $ 10.2$^{d}$  \\

&  Roots  & 130.21 $\pm $ 5.2$^{d }$ & 56.34 $\pm $ 4.1$^{e}$  \\

\hline
\end{tabular}
\end{center}
\end{table}

 Hypocotyl tTCLs were cultivated for 6 weeks on MS basal medium supplemented 
with NAA (3.22 $\mu$M), BAP (26.6 $\mu$M), Sucrose 2 \% (w/v), AgNO$_3$ 
(10 $\mu$M) and CdCl$_2$ (0 or 150 $\mu$M).

 The results were calculated from 3 independent experiments repeated 
3 times. For F.W., the values with different letters are the means $\pm$ S.D. 
and are significantly different at \textit{p} = 0.05 (ANOVA and LSD test).

\begin{table}[htbp]
\begin{center}
\caption{ Effect of CdCl$_2$ (75 $\mu$M) hydroponically applied for 3 days on 
the growth (D.W.) of plantlets regenrated from hypocotyl tTCL explants.} 
\label{tab3}
\begin{tabular}{ c c c c}
\hline  
CdCl$_{2}$ ($\mu $M) & & 0  & 150   \\
\hline

& Leaves  & 248.23 $\pm $ 6.51$^{a }$ & 128.24 $\pm $ 3.07$^{b}$  \\

0  & Stems &  42.80 $\pm $ 1.92$^{c }$ & 14.61 $\pm $ 0.17$^{e}$  \\

& Roots  & 27.91 $\pm $ 0.26$^{d }$ & 11.01 $\pm $ 0.10$^{e}$  \\

\hline

& Leaves  & 251.21 $\pm $ 11.81$^{a }$ & 129.27 $\pm $ 2.72$^{b}$  \\

75  & Stems &  43.31 $\pm $ 1.64$^{c }$ & 14.85 $\pm $ 0.35$^{e}$  \\

& Roots  & 27.86 $\pm $ 0.53$^{d }$ & 10.86 $\pm $ 0.09$^{e}$  \\

\hline
\end{tabular}
\end{center}
\end{table}

 Hypocotyl tTCLs were cultivated for 6 weeks on MS basal medium supplemented 
with NAA (3.22 $\mu$M), BAP (26.6 $\mu$M), Sucrose 2 \% (w/v), 
AgNO$_3$ (10 $\mu$M) and CdCl$_2$ (0 or 150 $\mu$M).

 The results were calculated from 3 independent experiments repeated 3 times. 
For F.W., the values with different letters are the means $\pm$ S.D. 
and are significantly different at $p$ = 0.05 (ANOVA and LSD test).

\begin{figure}[htbp]
\begin{center}
\scalebox{0.6}{\includegraphics*{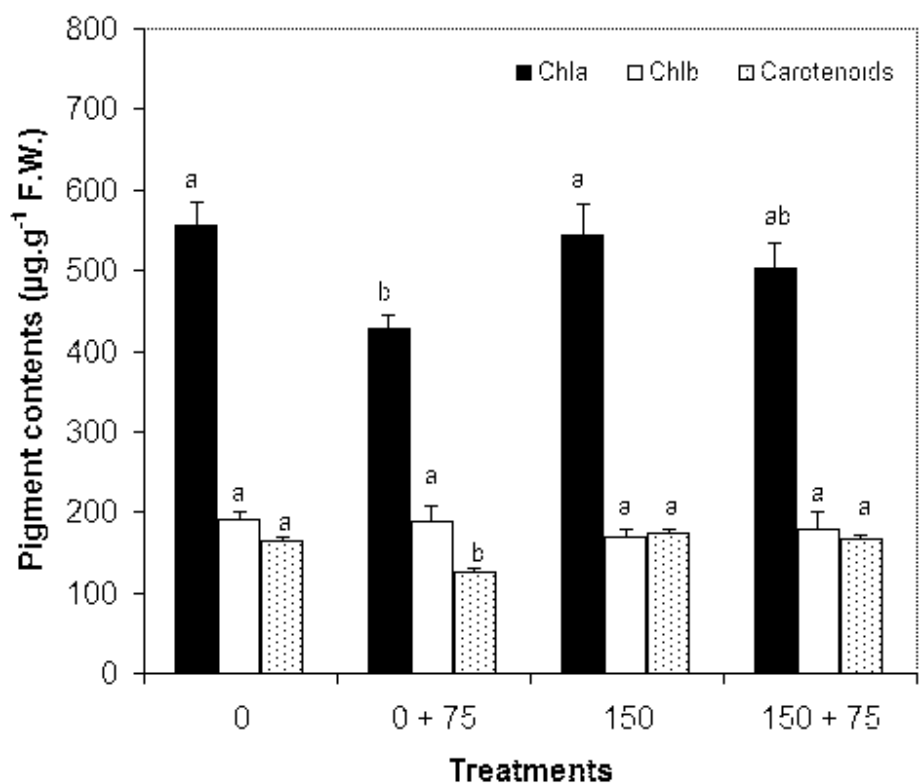}}
\end{center}
\caption{75 $\mu$M of CdCl$_2$ (hydroponically applied for 3 days) effect on pigment 
contents of Brassica juncea plantlets regenerated on MS basal medium 
supplemented with NAA (3.22 $\mu$M), BAP (26.6 $\mu$M), Sucrose 2 \% (w/v), AgNO$_3$ (10 
$\mu$M) and CdCl$_2$ (0 or 150 $\mu$M).
Results calculated from three independent experiments, repeated 3 times are 
means $\pm$ S.D. 
Different letters indicated significant difference at $p$ = 0.05 (One way ANOVA 
and LSD test).}
\label{fig1}
\end{figure}

\begin{figure}[htbp]
\begin{center}
\scalebox{0.6}{\includegraphics*{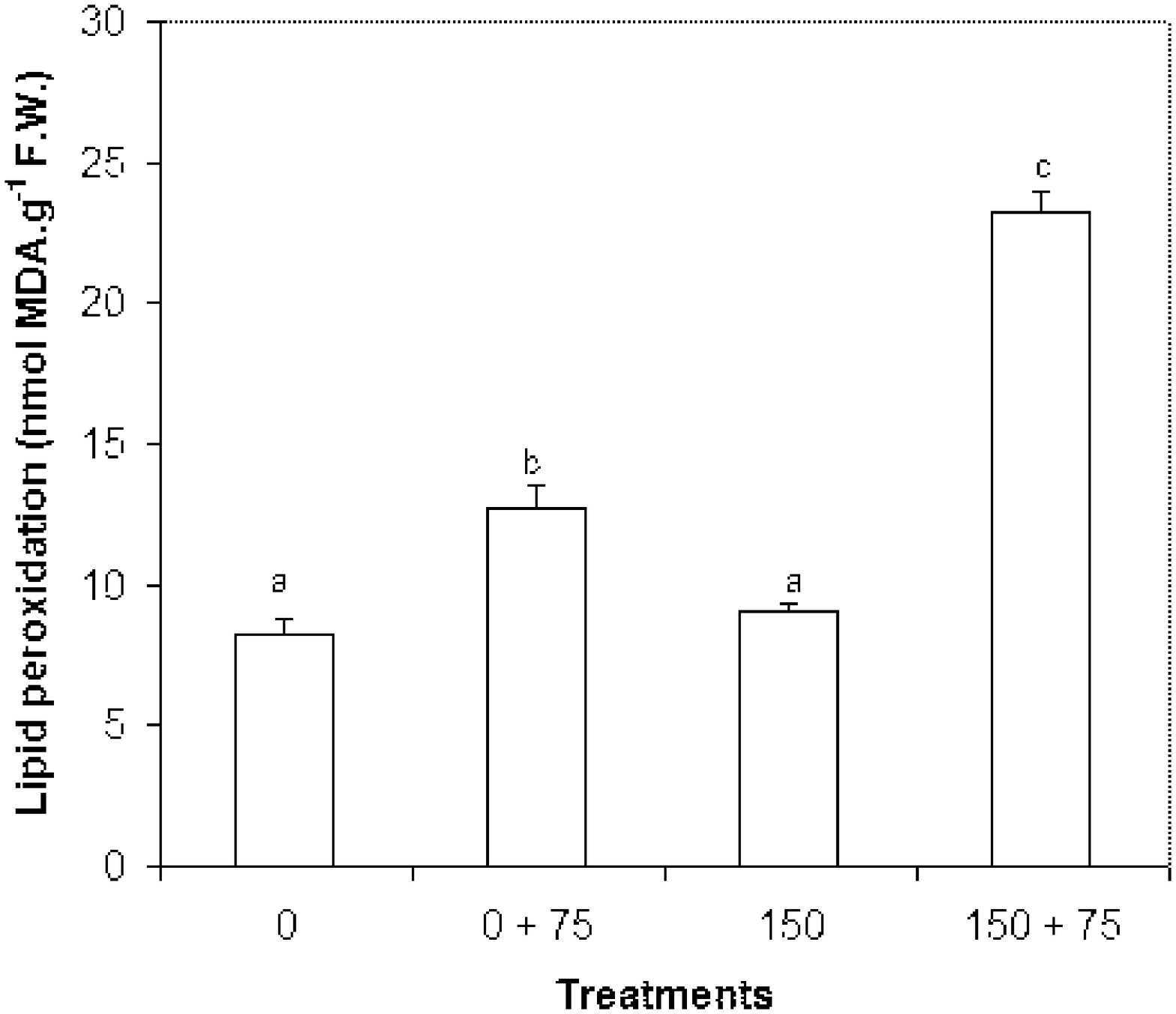}}
\end{center}
\caption{75 $\mu$M of CdCl$_2$ (hydroponically applied for 3 days) effect on MDA 
contents of Brassica juncea plantlets regenerated on MS basal medium 
supplemented with NAA (3.22 $\mu$M), BAP (26.6 $\mu$M), Sucrose 2 \% (w/v), AgNO$_3$ (10 
$\mu$M) and CdCl$_2$ (0 or 150 $\mu$M).
Results calculated from three independent experiments, repeated 3 times are 
means $\pm$ S.D. 
Different letters indicated significant difference at $p$ = 0.05 (One way ANOVA 
and LSD test).} 
\label{fig2}
\end{figure}

\end{document}